\documentclass[12pt,a4paper]{article}
\usepackage[utf8]{inputenc}
\usepackage{a4wide}
\usepackage{graphicx}
\usepackage{amsmath}
\usepackage{amssymb}
\usepackage{array}
\allowdisplaybreaks[1] 
\usepackage{hyperref}
\usepackage{cancel}
\usepackage{cite}
\usepackage{multirow}
\usepackage{slashed}
\usepackage{amssymb}
\usepackage{tabularx}
\usepackage{longtable}
\usepackage{xspace}

\usepackage[dvipsnames]{xcolor}

\usepackage{pifont}

\newcommand{\trace}[1]{\left\langle #1 \right\rangle}
\newcommand{\groupG}{\textrm{SU}(N_f) _\textrm{L}\times \textrm{SU}(N_f)_\textrm{R} }
\newcommand{\groupH}{\textrm{SU}(N_f)_\textrm{V} }

\newcommand{\ope}{\ensuremath{\mathcal{O}}\xspace}

\begin{document}

\newcolumntype{L}[1]{>{\raggedright\arraybackslash}p{#1}}
\newcolumntype{C}[1]{>{\centering\arraybackslash}p{#1}}
\newcolumntype{R}[1]{>{\raggedleft\arraybackslash}p{#1}}

\begin{titlepage}
\begin{flushright}
October 2023
\end{flushright}
\vfill
\begin{center}
{\Large\bf The anomalous chiral Lagrangian at order $p^8$
}
\vfill
{\bf Johan Bijnens $^{a,\, \dagger}$, Nils Hermansson-Truedsson $^{a,\, b, \, \ddagger}$, Joan Ruiz-Vidal $^{a, \, \ast}$}\\[0.3cm]
{$^{a}$ Division of Particle and Nuclear Physics, Department of Physics, Lund University,\\
Box 118, SE 221 00 Lund, Sweden
\\
$^{b}$ Higgs Centre for Theoretical Physics, School of Physics and Astronomy, \\The University of Edinburgh, James Clerk Maxwell Building,
Peter Guthrie Tait Road,
Edinburgh,
EH9 3FD
}
\end{center}
\vfill
\begin{abstract}
We derive the order $p^8$ Lagrangian of odd intrinsic parity for mesonic chiral perturbation theory, and provide the resulting operator basis in the supplementary material. Neglecting the non-zero singlet trace
, we find $999$ operators for a general number of quark flavours $N_f$, $705$ for $N_f=3$ and $92$ for $N_f=2$. Our numbers agree with those obtained through the Hilbert series approach in the literature. Including a singlet trace, as needed for the physical case of $N_f=2$, instead yields $1210$ operators for a general $N_f$, $892$ for $N_f=3$ and $211$ for $N_f=2$.
\end{abstract}
\vfill

{\footnotesize\noindent $^\dagger$ johan.bijnens@hep.lu.se
\\
\footnotesize\noindent $^\ddagger$ nils.hermansson-truedsson@ed.ac.uk
\\
\footnotesize\noindent $^\ast$ joan.ruiz-vidal@hep.lu.se
}
\end{titlepage}

\tableofcontents
\section{Overview}
\label{sec:introduction}
Chiral perturbation theory (ChPT) is the effective field theory of quantum chromodynamics (QCD) at low energies~\cite{Weinberg:1978kz,Gasser:1983yg,Gasser:1984gg}. The degrees of freedom are the $N_f^2-1$ lightest pseudoscalar mesons for QCD with $N_f$ light-quark flavours, which are the pseudo-Nambu-Goldstone bosons of the spontaneous chiral symmetry breaking $\groupG \rightarrow \groupH $. For $N_f=3$ these are the pions, kaons and the eta meson, whereas for $N_f=2$ only the pions. The power counting in ChPT is done in terms of a generic low-energy momentum transfer $p$, and the systematic ordering through $n$--loop order is $p^{2n}$. For a pedagogical introduction to ChPT, see e.g.~Ref.~\cite{Pich:2018ltt}. 

The effective construction of ChPT allows for systematically improvable predictions, order by order in the chiral counting. In the even intrinsic parity\footnote{ Intrinsic parity is parity without the space reflection, i.e. the (defined later) fields $\phi,a_\mu$ and $p$ go to minus themselves, $s$ and $v_\mu$ to themselves.} sector, current state-of-the art calculations are performed at order $p^8$~\cite{Bijnens:2017wba,Bijnens:2018lez,Hermansson-Truedsson:2020rtj}. The effective Lagrangians at orders $p^2$ and $p^4$ were first derived in Refs.~\cite{Gasser:1983yg,Gasser:1984gg}, order $p^6$ in Refs.~\cite{PhysRevD.53.315,Bijnens:1999sh,Weber:2008} and order $p^8$ in Ref.~\cite{Bijnens:2018lez}. At each order there are new low-energy constants (LECs), whose renormalized values are needed for ChPT predictions. Renormalization of the LECs is known through order $p^6$, see Refs.~\cite{Gasser:1983yg,Gasser:1984gg,Bijnens:1999hw}. 

The Lagrangian for odd intrinsic parity processes starts at order $p^4$, governed by the Wess-Zumino-Witten anomaly~\cite{WESS197195,Witten:1983tw}, and allows processes such as $\pi^0 \rightarrow \gamma \gamma $. Although the anomaly itself is captured by the Wess-Zumino-Witten term, there are higher-order anomalous terms in the ChPT Lagrangian. In Ref.~\cite{Bijnens:2001bb,Ebertshauser:2001nj}, the order $p^6$ anomalous Lagrangian was determined. The associated renormalization of LECs was determined in Refs.~\cite{Bijnens:1989jb,Akhoury:1990px,Issler:1990nj,Bijnens:2001bb}. Reviews of this sector are Refs.~\cite{Bijnens:1993xi,Weber:2008}.

Effective Lagrangians are in general constructed from a set of building blocks combined in all possible ways into so-called monomials that satisfy the underlying symmetries of the theory. With ChPT as an effective field theory of QCD, the relevant ones are chiral symmetry (from QCD with $N_f$ massless quarks), parity ($\mathcal{P}$), charge ($\mathcal{C}$) and hermitian conjugation (h.c.)~\cite{Gasser:1983yg,Gasser:1984gg}. Monomials are often related through certain relations, e.g.~integration-by-parts identities. A naive construction of a monomial basis satisfying the imposed symmetries will thus be over-complete, and it is of interest to minimise the number of monomials. The minimal number of terms can be found in several ways, either by the Hilbert series approach~\cite{Henning:2017fpj,Graf:2020yxt,Bijnens:2022zqo} or by a systematic use of the monomial relations~\cite{Bijnens:1999sh,Bijnens:2001bb,Weber:2008,Bijnens:2018lez,Bijnens:2022zqo}. Only the latter method gives an explicit expression for the Lagrangian. For a comparison between the two approaches for the $\textrm{O}(N)$ non-linear sigma model, see Ref.~\cite{Bijnens:2022zqo}. For the construction of a monomial basis, one can use the systematic approach mentioned above or one based on the Adler zero, see Refs.~\cite{Low:2014nga,Low:2022iim} and references therein. 

In this paper we derive the order $p^8$ chiral Lagrangian in the odd intrinsic parity sector, by systematically enumerating all monomials allowed by symmetry and using relations to minimise the corresponding set of terms. The minimal number of terms was obtained already in Ref.~\cite{Graf:2020yxt}, but using an independent method based on the Hilbert series. We confirm and extend their results, and in addition construct for the first time an explicit basis of the effective Lagrangian. Due to the length of the expression, the basis is presented in the supplementary material.

The calculation, as outlined in the paper, has been independently implemented in two software platforms: {\sc{Form}}~\cite{Vermaseren:2000nd,Ruijl:2017dtg} and \textsc{Mathematica}~\cite{Mathematica}. The \textsc{Mathematica} version of the code is publicly accessible as part of the {\tt MINIBAR} package~\cite{MINIBAR}, which offers a range of tools to help the extraction of minimal operator bases in a time-efficient way.

The structure of this manuscript is as follows. In section~\ref{sec:buildingblocks} we introduce the chiral building blocks and the operator basis we use for the Lagrangian. The various types of relations between monomials constructed from the operators are presented in section~\ref{sec:relations}, and in section~\ref{sec:lagresult} we present our result. Conclusions and an outlook are given in section~\ref{sec:conclusion}.

\section{The chiral Lagrangian}\label{sec:buildingblocks}

In this section, we present the context of our work, chiral perturbation theory, and outline the systematic organization of the chiral Lagrangian. In preparation to construct this Lagrangian, the definition of the building blocks to compose chiral-invariant operators is provided, as well as the method to combine these into symmetric monomials\footnote{%
	Throughout this manuscript, we use the term "operator" to refer to a basic field interaction term invariant under chiral symmetry and composed of the building blocks, while the term "monomial" denotes linear combinations of operators that are either even or odd under $\mathcal{P}$, $\mathcal{C}$ and hermitian conjugation.
	
}. It is the monomials that enter the effective Lagrangian, each multiplied by a LEC.

\subsection{A brief introduction to chiral perturbation theory}
ChPT is the effective field theory of QCD at low energies~\cite{Weinberg:1978kz,Gasser:1983yg,Gasser:1984gg}, constructed from the approximate chiral symmetry $G= \groupG$ for $N_f$ light quarks. The spontaneous breaking of the chiral symmetry $\groupG\rightarrow \groupH $ is due to a non-zero quark condensate $\langle \overline{q}q\rangle \neq 0$, where $q$ is a column-vector of the $N_f$ light-quark fields. The pseudo-Nambu-Goldstone bosons, i.e.~the $N_f^2-1$ lightest pseudoscalar mesons contained in the matrix $\phi$, live in the coset space $G/H$ where $H=\groupH$.  

To account for quark masses and interactions beyond pseudoscalars, external fields can be added to the underlying QCD Lagrangian $\mathcal{L}^0_{\textrm{QCD}}$~\cite{Gasser:1983yg,Gasser:1984gg}. Denoting the scalar, pseudoscalar, vector and axial vector external fields by $N_f\times N_f$ flavour matrices $s$, $p$, $v_\mu$ and $a_\mu$, respectively, the modified QCD Lagrangian can be written
\begin{align}
\mathcal{L} = \mathcal{L}^0_{\textrm{QCD}}+
\overline q\gamma^\mu\left(v_\mu+a_\mu\gamma_5\right)-\overline q \left(s-ip\gamma_5\right)q \, .
\end{align}
The external fields transform such that the chiral symmetry $G$ can be promoted to a local symmetry of $\mathcal{L}$, and consequently ChPT is constructed with $G$ local and external fields included. For later use we introduce the combinations
\begin{align}
\label{defchi}
    \chi & \equiv  2B\left(s+ip\right) \, ,
    \\
    \ell_\mu & \equiv v_\mu-a_\mu \, ,
    \\
    r_\mu & \equiv v_\mu+a_\mu \, .
\end{align}
The definition of $\chi$ traditionally includes a LEC $B$ \cite{Gasser:1983yg,Gasser:1984gg} which is related to $\langle \overline{q}q\rangle$.

The building blocks in the chiral Lagrangian depend on the meson fields $\phi$ as well as external fields $\chi$, $\ell _\mu$ and $r_\mu$. In the CCWZ formalism~\cite{PhysRev.177.2239,PhysRev.177.2247}, the meson fields can be collected into the unitary $N_f\times N_f$ matrix $u(\phi)$. Under chiral transformations in terms of group element $(g_L,g_R)\in G$ the fields transform according to 
\begin{align}
\label{eq:transformation}
u(\phi) &\,\longrightarrow g_R u(\phi) h(g_L,g_R, u(\phi))^\dagger = h(g_L,g_R, u(\phi))u(\phi)g_L^\dagger\,,
\nonumber\\
\chi &\,\longrightarrow g_R\chi g_L^\dagger\,,
 \nonumber\\
\ell_\mu &\,\longrightarrow g_L \ell_\mu g_L^\dagger-i\partial_\mu g_L g_L^\dagger\,,
\nonumber\\
r_\mu &\,\longrightarrow g_R \ell_\mu g_R^\dagger-i\partial_\mu g_R g_R^\dagger\,.
\end{align}
Here we see the group element $h(g_L,g_R,u(\phi))\in H$, the so-called compensator field, which depends in particular on $\phi$. For convenience we will not write the arguments of $u$ and $h$.

Anomalies arise when the symmetries of the Lagrangian at the classical level are not respected in the quantized theory. This is the case of QCD, whose Lagrangian features a general axial symmetry that is nevertheless violated after cancelling the divergences in processes with fermion-loop triangle diagrams~\cite{Bell:1969ts,Adler:1969gk}. The effective theory of QCD, ChPT, captures the anomaly in the WZW interaction term~\cite{WESS197195,Witten:1983tw}, which is not invariant under $G$ already at the classical level. This is also the leading interaction involving an odd number of Goldstone bosons. This feature is commonly referred to as odd \textit{intrinsic} parity, as it is equivalent to normal $\mathcal{P}$ transformations but ignoring the space-time coordinates, \textit{e.g.} $\phi(\vec{x},t) \to -\phi(\vec{x},t)$ for pseudoscalar meson fields.

The effective mesonic chiral Lagrangian is constructed systematically in powers of $p^2$, where $p$ is the typical momentum scale in a low-energy process. To this end we may write
\begin{align}
\mathcal{L} _{\chi} = \mathcal{L}^{\textrm{even}} + \mathcal{L}^{\textrm{odd}} \, . 
\end{align}
The Lagrangian $\mathcal{L}^{\textrm{even}}$ only contains terms of even intrinsic parity, and starts at order $p^2$. It is currently known through order $p^8$~\cite{Gasser:1983yg,Gasser:1984gg,PhysRevD.53.315,Bijnens:1999sh,Weber:2008,Bijnens:2018lez} and it is not the subject of this paper. Instead, we are interested in the so-called anomalous Lagrangian $\mathcal{L}^{\textrm{odd}}$, of odd intrinsic parity, which starts at order $p^4$ and, previous to this publication, was known up to order $p^6$~\cite{WESS197195,Witten:1983tw,Bijnens:2001bb,Ebertshauser:2001nj}.

\subsection{The anomalous chiral Lagrangian}

Unlike the non-anomalous sector, $\mathcal{L}^{\textrm{odd}} $ starts at order $p^4$. We may thus write 
\begin{align}
    \mathcal{L}^{\textrm{odd}} = \mathcal{L}_4^{\textrm{odd}} + \mathcal{L}_6^{\textrm{odd}} + \mathcal{L}_8^{\textrm{odd}} +\ldots \, .
\end{align}
The leading order $\mathcal{L}_4^{\textrm{odd}}$ is given by the Wess-Zumino-Witten term~\cite{WESS197195,Witten:1983tw} and it is completely fixed by one parameter that can be identified with the number of colours $N_c$ in QCD. This is also the only term containing non-invariant structures under chiral symmetry. Beyond $\mathcal{L}_4^{\textrm{odd}}$, new effective operators are needed to renormalize the singularities arising in loop diagrams with $\mathcal{L}_4^{\textrm{odd}}$ vertices, as well as to provide improved effective descriptions of higher order interactions from QCD. The next-to-leading order Lagrangian $\mathcal{L}_6^{\textrm{odd}}$ was derived in Refs.~\cite{Bijnens:2001bb,Ebertshauser:2001nj} and contains several additional parameters.
The purpose of this work is to provide the next-to-next-to-leading order Lagrangian $\mathcal{L}_8^{\textrm{odd}}$.
Starting from order $p^6$, the effective Lagrangians take the form
\begin{align}\label{eq:lagdef2n}
 \mathcal{L}_{2n}^{\textrm{odd}} = 
 \sum _{i=3}^{N_{2n}} \mathcal{O}_i^{(2n)} \, a_{i}^{(2n)} \, ,
\end{align}
where $\mathcal{O}_i^{(2n)}$ is a Lorentz-invariant monomial of chiral order $p^{2n}$ invariant under chiral symmetry, parity, charge and hermitian conjugation. The monomials consist of (products of) flavour traces of building blocks, containing the meson and external fields, which are $N_f\times N_f$-matrices in flavour space~\cite{Gasser:1983yg,Gasser:1984gg}. The $a_{i}^{(2n)}$ in Eq.~(\ref{eq:lagdef2n}) are LECs and $N_{2n}$ is the number of monomials in a minimal basis depending on the number of light-quark flavours, $N_f$. To preserve the overall parity of the odd-intrinsic-parity monomials in $\mathcal{L}^{\textrm{odd}}$, every monomial contains a Levi-Civita tensor $\epsilon ^{\mu \nu \rho \sigma }$ such that 
\begin{align}\label{eq:anomon}
    \mathcal{O}_i^{(2n)} = \mathcal{O}_{i, \, \mu  \nu  \rho \sigma }^{(2n)} \epsilon ^{\mu  \nu \rho \sigma } \, .
\end{align}

Among these monomials, those that do not contain Goldstone bosons are not measurable and we identify them separately in our Lagrangian basis. They can be constructed purely in terms of the external fields $s,p,\ell_\mu,r_\mu$ and are called contact operators. To make these external-field contact operators explicit in the final Lagrangian, they are constructed separately. In the following sections we introduce the two field parametrizations that form the building blocks for the monomials in the Lagrangian, the first one more appropriate for the contact operators, the second for all operators.

\subsection{The left-right basis}

In ChPT, the meson and external fields are parametrized through matrices in flavour space that transform linearly under chiral transformations. The left-right (LR) basis is a broadly used parametrization in which both left and right matrices of the chiral group, $g_L,g_R \in SU(3)_L \times SU(3)_R$, appear in the transformation of the fields as

\begin{align}\label{eq:defU}
U\equiv u^2&\longrightarrow  g_{R}  U  g_{L}^\dagger \,,
\nonumber\\
\chi&\longrightarrow  g_{R}  \chi  g_{L}^\dagger \,,
\nonumber\\
 F_{L}^{\mu \nu} \equiv \partial ^{\mu}\ell ^{\nu}-\partial ^{\nu}\ell ^{\mu} -i\left[ \ell ^{\mu},\ell ^{\nu} \right]&\,\longrightarrow
g_L  F_{L}^{\mu \nu} g_L^\dagger\,,
\nonumber\\
 F_{R}^{\mu \nu} \equiv \partial ^{\mu}r ^{\nu}-\partial ^{\nu}r ^{\mu} -i\left[ r ^{\mu},r ^{\nu} \right] &\,\longrightarrow
g_R  F_{R}^{\mu \nu} g_R^\dagger\,.
\end{align}
In this basis, the $U$ matrix contains the Goldstone bosons, $\chi$ the external scalar and pseudoscalar fields, and $F_{L(R)}^{\mu \nu }$ is the field-strength tensor of the left (right) external sources cf.~Eq.~\eqref{eq:transformation}. Depending on the chiral transformation properties of the matrix $O$, its covariant derivative is defined as
\begin{align}\label{eq:covder}
D_{\mu} O =
\left\{
\begin{array}{ll}
 \partial _{\mu}O -ir_{\mu}O +iO \ell _{\mu},  & O\longrightarrow g_{R}\, O\, g_{L}^{\dagger}  
\, ,  \\ \\ 
\partial _{\mu}O  -i \ell _{\mu}O+iOr_{\mu},  & O\longrightarrow g_{L}\, O\, g_{R}^{\dagger}  
\, ,  \\ \\
\partial _{\mu}O -ir_{\mu}O +i O r _{\mu},  & O\longrightarrow g_{R}\, O\, g_{R}^{\dagger}  
\, ,  \\ \\
\partial _{\mu}O -i \ell_{\mu}O +i O\ell _{\mu},  & O\longrightarrow g_{L}\, O\, g_{L}^{\dagger}  
\, .  \\
\end{array}
\right. 
\end{align}
The transformation properties of these building blocks under the discrete symmetries parity, charge conjugation and hermitian conjugation are given in Table \ref{table:disctransLR}. In the chiral counting the building blocks scale as $U\sim 1$, $D_\mu \sim p$, $\chi \sim p^2$ and $F_{L,R}^{\mu \nu}\sim p^2$. 

To construct all possible operators invariant under the symmetries of the theory, the building blocks in Eq.~\eqref{eq:defU} must be combined in all possible ways, including all permutations of indices, and distributed over different traces.  Nevertheless, chiral-invariant structures proportional to $\det \chi$ cannot be generated using only combinations of building blocks. To integrate these additional invariant structures in the method, the building block $\tilde\chi$ of chiral order $p^{2 N_f}$ is introduced, which transforms as $\chi$~\cite{Gasser:1983yg,Kaplan:1986ru,Bijnens:2018lez},
	\begin{align}
	\label{eq:chit}
	\tilde\chi\equiv\left(\det(\chi)\chi^{-1}\right)^\dagger
	\longrightarrow g_R \tilde\chi g_L^\dagger\,.
	\end{align}
	The operators with $\tilde\chi^{(\dagger)}$ are clearly not independent of those with $\chi^{(\dagger)}$. To systematically verify these dependencies when employing the LR basis, the traces with $\tilde\chi$ can be written in terms of the matrix elements of $\chi$, with the explicit expression for $\tilde\chi$ given in Ref.~\cite{Bijnens:2018lez} and below in Eq.~\eqref{eq:chitilde}.
	
\begin{table}
	\begin{center}  
		\renewcommand{\arraystretch}{1.2}
		\begin{tabular}{|C{2cm}|C{3cm}|C{3cm}|C{3cm}|}
			\hline
			& $P$ & $C$ & h.c. \\ \hline 
			$U$         & $U ^{\dagger}$ &  $U ^{T}$& $U ^{\dagger}$ \\
			$\chi$         & $\chi ^{\dagger}$ &  $\chi ^{T}$& $\chi ^{\dagger}$ \\
			$F_{L}^{\mu \nu }$     & $\varepsilon(\mu)\varepsilon(\nu)F_{R}^{\mu \nu}$ & $-\big( F_{R}^{\mu \nu}\big) ^{T}$ & $F_{L}^{\mu \nu}$ \\
			$F_{R}^{\mu \nu}$     & $\varepsilon(\mu)\varepsilon(\nu)F_{L}^{\mu \nu }$ & $-\big( F_{L}^{\mu \nu}\big) ^{T}$& $F_{R}^{\mu \nu}$ \\
			\hline
		\end{tabular}
		\renewcommand{\arraystretch}{1}
		\caption{Discrete transformation properties of the chiral building blocks in the LR basis. Here we use the convention $\varepsilon(0)=-\varepsilon(i=1,2,3)=1$.}\label{table:disctransLR}
	\end{center}
\end{table}

\subsection{The $u$ basis}

While the LR basis is broadly used in phenomenology due to its relatively direct connection to meson fields, introducing an additional step in the parametrization of these fields offers distinct advantages. In particular, it simplifies certain calculations and facilitates a more straightforward connection to group-theoretical methods.

In the $u$ basis, introduced in Ref.~\cite{Ecker:1988te}, each building block $X$ transforms only under the vector subgroup $H$ as $X\rightarrow h Xh^\dagger$, automatically keeping the chiral invariance under permutations of the building blocks. This basis is particularly convenient for building effective Lagrangians as it puts the building blocks on equal footing and simplifies the extraction of operator relations, described in section~\ref{sec:relations}. The new building blocks are related to the LR-basis through
\begin{align}
\label{eq:defblocks}
 u_{\mu}&\, = i\Big[ u^{\dagger} (\partial _{\mu}-ir_{\mu})u-u(\partial _{\mu}-i\ell _{\mu} )u^{\dagger}\Big]
\, ,\nonumber \\
 \chi _{\pm}&\, = u^{\dagger}\chi u^{\dagger}\pm u\chi ^{\dagger} u
\, , \nonumber\\
 f_{\pm}^{\mu \nu}&\, = u F_{L}^{\mu \nu }u^{\dagger} \pm u^{\dagger}F_{R}^{\mu \nu}u
 \, .
\end{align}
Another advantage of the $u$ basis is that all buildings blocks share the same covariant derivative, unlike in Eq.~\eqref{eq:covder}. This is defined as
\begin{align}\label{eq:nablader}
& \nabla _{\mu} X= \partial _{\mu}X +\left[ \Gamma _{\mu}, X\right] 
\, ,
\end{align}
in which $\Gamma_\mu$ is the chiral connection
\begin{align}
\Gamma _{\mu} = \frac{1}{2}\Big[ u^{\dagger}\left( \partial _{\mu}-ir_{\mu}\right) u+u\left( \partial _{\mu}-i\ell _{\mu}\right) u^{\dagger} \Big] \, .
\end{align}

The transformation properties of this basis under the discrete symmetries is given in Table \ref{table:disctrans}.
Note that in the LR basis the meson fields are contained solely in the $U$ matrices whereas in our basis both $u_\mu$, $f_{\pm \mu \nu}$ and $\chi_\pm$ all contribute to processes with Goldstone bosons. Intrinsic parity is easier in the $u$-basis since $u_\mu$, $f_{-\mu\nu}$ and $\chi_-$ only change sign. So odd intrinsic parity implies an odd number of these fields in the operators.

\begin{table}
  \begin{center}  
  \renewcommand{\arraystretch}{1.5}
    \begin{tabular}{|C{2cm}|C{3cm}|C{3cm}|C{3cm}|}
      \hline
                        & $P$ & $C$ & h.c.  \\ \hline 
      $u_{\mu}$         & $-\varepsilon (\mu) u_{\mu}$ &  $u_{\mu}^{T}$& $u_{\mu}$ \\
      $\chi _{\pm}$     & $\pm \chi _{\pm}$ & $\chi _{\pm}^{T}$& $\pm \chi _{\pm}$ \\
      $f_{\pm \mu \nu}$ &  $\pm\varepsilon (\mu)\varepsilon (\nu) f_{\pm \mu \nu}$ & $\mp f_{\pm \mu \nu}^{T}$ & $f_{\pm \mu \nu}$ \\ \hline
    \end{tabular}
   \renewcommand{\arraystretch}{1}
     
    \caption{Discrete transformation properties of the chiral building blocks in our basis. Here we have the convention $\varepsilon (0) = -\varepsilon (i=1,2,3) = 1$.}\label{table:disctrans}
  \end{center}
\end{table}

\subsection{Translation between bases} \label{sec:translation}

To include the contact terms, written in the LR basis, as part of the main Lagrangian, in the $u$ basis, their independence to other operators must be ensured. The external field matrices can be translated using 
\begin{align}
	\label{eq:translationFields}
	\chi &\,= \frac{1}{2}u\left( \chi _{+}+\chi _{-}\right) u 
	\, , \nonumber\\
	\chi^{\dagger}&\, = \frac{1}{2}u^{\dagger}\left( \chi _{+}-\chi _{-}\right) u^{\dagger} 
	\, ,\nonumber \\
	F_L^{\mu\nu}&\, = \frac{1}{2}u^{\dagger} \left( f_{+}^{\mu \nu}+f_{-}^{\mu \nu}\right)u
	\, ,\nonumber \\ 
	F_R^{\mu \nu}&\, = \frac{1}{2}u \left( f_{+}^{\mu \nu}-f_{-}^{\mu \nu}\right)u^\dagger \, .
	\end{align}
	The translation of $\tilde\chi^{(\dagger)}$, only needed here for $N_f=2$, can be derived from the definition in Eq.~(\ref{eq:chit}) and the relation between determinant and trace of $2\times 2$ matrices. It reads
	\begin{align}
	\label{eq:translationChit}
	\trace{A \tilde\chi} = \trace{u A u } \trace{u \chi^\dagger u}  -  \trace{u A u u \chi^\dagger u} \, ,
	\end{align}
where $A$ is any matrix transforming as $A\longrightarrow g_L A g_R$. The analoguous formula for $\tilde\chi^\dagger$ is the hermitian conjugate of Eq.~\eqref{eq:translationChit}.

As seen in Eq.~\eqref{eq:covder}, the covariant derivative in the LR basis is defined differently depending on the transformation properties of the matrix on which it acts. Consequently, the relation between the LR covariant derivatives $D_\mu$ and the $\nabla_{\mu }$ derivative of the $u$ basis, defined in Eq.~\eqref{eq:nablader}, also depends on these transformation properties and is given by 
\begin{align}
\label{eq:translationDerivatives}
~ u^\dagger D_\mu O u^\dagger &\,= \nabla_\mu (u^\dagger O u^\dagger) - \frac{i}{2}\left\lbrace u_\mu , u^\dagger O u^\dagger \right\rbrace &\text{for}~~ O \longrightarrow g_R O g_L^\dagger
\, , \nonumber\\
~ u D_\mu O u &\,= \nabla_\mu (u O u) + \frac{i}{2}\left\lbrace u_\mu , u O u \right\rbrace&  \text{for}~~ O \longrightarrow g_L O g_R^\dagger
\, , \nonumber\\
~ u^\dagger D_\mu O u &\,= \nabla_\mu (u^\dagger O u) - \frac{i}{2}\left[ u_\mu , u^\dagger O u \right]&  \text{for}~~ O \longrightarrow g_R O g_R^\dagger
\, , \nonumber\\
~ u D_\mu O u^\dagger &\,= \nabla_\mu (u O u^\dagger) + \frac{i}{2}\left[  u_\mu , u O u^\dagger \right]&   \text{for}~~ O \longrightarrow g_L O g_L^\dagger
\, .
\end{align}

\subsection{Enforcing the symmetries}\label{sec:discrete}

Besides being chirally invariant, the ChPT Lagrangian must also be symmetric under Hermitian conjugation, parity and charge conjugation.
The simplest way to make sure that the discrete symmetries of the Lagrangian are respected is to symmetrize each monomial individually in the over-complete basis. By writing each operator as a linear combination with its transformed counterpart, the resulting monomials become eigenvectors of the symmetry. To enforce hermiticity, the operators in the initial basis $\lbrace \ope_i \rbrace$ are transformed as
\begin{align}
\label{eq:symH}
\tilde\ope_i &\,= \ope_i + i \ope_i^\dagger \, , & \, \tilde\ope'_i &= \ope_i - i \ope_i^\dagger \, .
\end{align}
It may seem that this method doubles the number of monomials in the $\lbrace \tilde\ope_i \rbrace$ basis, but because both $\ope_i$ and $\ope_i^\dagger$ exist in the original $\lbrace \ope_i \rbrace$ basis, the monomials generated by $\ope_i^\dagger$ are always proportional to those generated by $\ope_i$, and then they can be identified as redundant and be removed. In the case of hermitian conjugation, the resulting odd terms can be made even by simply adding an overall $i$ factor. The resulting basis is then symmetrized under parity and, subsequently, under charge conjugation. In these cases, the transformations are
\begin{align}
\label{eq:symPC}
\tilde\ope_i &\,= \ope_i +  \mathcal{P}(\ope_i) \, , & \, \tilde\ope'_i &= \ope_i -  \mathcal{P}(\ope_i)
\, , \nonumber\\
\tilde\ope_i &\,= \ope_i +  \mathcal{C}(\ope_i) \, , & \, \tilde\ope'_i &= \ope_i -  \mathcal{C}(\ope_i)
\, , \nonumber
\end{align}
where the transformation properties of the chiral matrices were given in Tables~\ref{table:disctransLR}--\ref{table:disctrans}. Note that in the $u$ basis all building blocks transform under $\mathcal{P}$ to themselves up to a possible sign, which means that in the initial construction of the monomials in the over-complete basis one can very easily isolate those that are even under parity. Once these are chosen in the $u$ basis, we thus only need to symmetrize under $\mathcal{C}$ explicitly. This does not, however, work in the LR basis since the building blocks transform in a non-trivial way. For the contact terms which are generated in the LR basis, we therefore have to symmetrize under both $\mathcal{P}$ and $\mathcal{C}$.
The main advantage of this approach to symmetrize the basis lies in its systematic and uniform application to all elements. An equivalent but technically more complex approach was used in Ref.~\cite{Bijnens:2018lez}.

\section{Relations to obtain a minimal basis}\label{sec:relations}
When systematically constructing all possible monomials for effective Lagrangians such as that in Eq.~(\ref{eq:lagdef2n}), one generally gets an over-complete basis with $N^0_{2n}>N_{2n}$ monomials $\mathcal{O}_j^{(2n)}$. In order to minimise the basis one must list all possible relations between ${\mathcal{O}}_j^{(2n)}$ following from a set of identities, namely
\begin{itemize}
    \item Partial integration or invariance of the action from total derivatives,
    \item Equations of motion or, equivalently, field redefinitions,
    \item Commuting of derivatives,
    \item The Bianchi identity,
    \item The Schouten identity,
    \item Additional relation with $f_{-}^{\mu \nu}$,
    \item The Cayley-Hamilton theorem.
\end{itemize}
In the following subsections we introduce each of the above classes of relations/identities. Since the difference between monomials and operators is only coming from the linear combinations under parity, charge and hermitian conjugation described in section~\ref{sec:discrete}, relations also hold at the operator level. For this reason we only consider operator relations below. 

\subsection{Partial integration relations}
The action is invariant under the addition of a total derivative in the Lagrangian. To include all such possibilities in the construction of the operators, we first construct all possible anomalous combinations of the building blocks at chiral order $p^{2n-1}$. Taking the derivative of these combinations thus produces a relation between operators at order $p^{2n}$. As an example, one obtains the relation\footnote{The $=0$ is to be understood as not contributing to the action.}
\begin{align}
0 
& =
\partial ^{\alpha } \trace{\chi _{-}f_{+ \alpha \mu } f_{- \nu \rho} u_{\sigma }} \, \epsilon ^{\mu  \nu \rho \sigma}\, .
\end{align}
By bringing the derivative inside the trace we obtain
\begin{align}
0& = 
\Big[ \trace{ \nabla ^{\alpha} \chi _{-}f_{+ \alpha \mu } f_{- \nu \rho} u_{\sigma}} 
+ \trace{  \chi _{-}\nabla ^{\alpha }f_{+ \alpha \mu } f_{- \nu \rho} u_{\sigma}}
\nonumber \\
& 
+ \trace{  \chi _{-} f_{+ \alpha \mu } \nabla ^{\alpha} f_{- \nu \rho} u_{\sigma}}
+ \trace{  \chi _{-} f_{+ \alpha \mu }  f_{- \nu \rho} \nabla ^{\alpha} u_{\sigma}} \Big]
\, \epsilon ^{\mu \nu \rho \sigma}
\, .
\end{align}
Here one may now recognise each of the four terms as an operator in the over-complete basis.

\subsection{Equation of motion relations}
The leading-order equation of motion of the chiral Lagrangian takes the form
\begin{align}\label{eq:eom}
\nabla ^{\mu}u_{\mu} -\frac{i}{2}\Big( \chi _{-}-\frac{1}{N_{f}}\langle \chi _{-}\rangle  \Big) =0 \, .
\end{align}
This equation can be used to generate relations between the operators in the over-complete basis, since combinations of operators that vanish using the leading-order equations of motion can be removed using field redefinitions of the building blocks~\cite{Scherer:1994wi,Bijnens:1999sh}. We can systematically generate all possible relations from the equation of motion by substituting the left-hand side of Eq.~\eqref{eq:eom} into the instances of $\nabla^{\mu}u_{\mu}$ within each operator, addressing one occurrence at a time. An equivalent approach to finding operator relations involves reconstructing all invariant structures, including an additional building block of order $p^2$, and then replacing this with Eq.\eqref{eq:eom}. An example relation is
\begin{align}
 0 
 & =
 \Big[ 
 \trace{ \nabla ^{\beta} u_{\beta} \nabla _{\mu } \nabla _{\nu} u _{\rho}}
 -\frac{i}{2}  \trace{ \chi _{-}\nabla _{\mu } \nabla _{\nu} u _{\rho}}
 \nonumber \\
 &
 + \frac{i}{2\, N_f}  \trace{ \chi _{-}}\trace{\nabla _{\mu } \nabla _{\nu} u _{\rho}}
 \Big]
 \, \trace{u^{\alpha } u_{\alpha} u_{\sigma} } \,  \epsilon ^{\mu \nu \rho \sigma} \, .
\end{align}

\subsection{The commutation of derivatives and the Bianchi identity}

The field-strength tensor $\Gamma _{\mu \nu}$ is given by 
\begin{align}
\label{eq:Gammamunu}
\Gamma_{\mu\nu} = \partial_\mu\Gamma_\nu-\partial_\nu\Gamma_\mu+\left[\Gamma_\mu,\Gamma_\nu\right]=\frac{1}{4}\left[ u_{\mu}, u_{\nu}\right] -\frac{i}{2}f_{+\mu \nu}\,.
\end{align}
This enters the relation
\begin{align}
\label{eq:fsdef}
\left[ \nabla _{\mu},\nabla _{\nu} \right] X = \left[ \Gamma _{\mu \nu},X\right]  \, ,
\end{align}
which we refer to as the commuting derivative relation. We generate all possible relations by once replacing each occurence of $\nabla _{\mu} \nabla_\nu X$ in each $\mathcal{O}^{(8)}_i$ in the over-complete basis. 

The field-strength tensor $\Gamma_{\mu\nu}$ further satisfies the so-called Bianchi identity
\begin{align}
B_{\mu\nu\rho}\equiv\nabla _{\mu} \Gamma _{\nu \rho} +\nabla _{\nu} \Gamma _{\rho \mu}+\nabla _{\rho} \Gamma _{\mu \nu} = 0 \, .
\end{align}
From the definition in Eq.~(\ref{eq:Gammamunu}) we immediately obtain the Bianchi relation
\begin{align}
\label{eq:defBBianchi}
B_{\mu\nu\rho} = \frac{1}{4}\Big( \big[ u_{\rho},f_{-\mu \nu}\big]  +\big[ u_{\mu},f_{-\nu \rho}\big]+ \big[ u_{\nu},f_{-\rho \mu}\big] \Big) -\frac{i}{2}\Big(  \nabla _{\rho} f_{+\mu \nu} + \nabla _{\mu}f_{+\nu \rho} + \nabla _{\nu} f_{+\rho \mu }\Big) =0 \, .
\end{align}
Note that the combination $B_{\mu\nu\rho}$ is of chiral order $p^3$, and has positive intrinsic parity. To generate all possible operator relations from the Bianchi identity we can again follow two different approaches: inserting the relation in the operators of the over-complete basis or generating a new list of anomalous combinations of building blocks at order $p^8$ using $B_{\mu\nu\rho}$ as an additional building block, which is allowed to occur only once per term.

\subsection{The Schouten identity}
The Schouten identity~\cite{Schouten:1938} asserts that in $d$ dimensions there can be no completely anti-symmetric tensors with more than $d$ Lorentz indices. For $N>d$, the identity can be written
\begin{align}\label{eq:schoutenid}
    0= \sum _{\{\sigma (\rho_1 \ldots \rho _N)\}} \pm\mathcal{O} _{\sigma(\rho_1 \ldots \rho _N)} \, ,
\end{align}
for an operator\footnote{We emphasise again that the identity has to be applied to any set of 5 indices in a given operator.} $\mathcal{O} _{\rho_1 \ldots \rho _N}$ and $\sigma (\rho_1 \ldots \rho _N)$ a permutation of the Lorentz-indices $\rho_1 \ldots \rho _N$ with the sign $\pm$ for an even or odd permutation. The sum runs over all permutations of the $N$ indices. To simplify notation, we have here only written out the $N$ indices of interest in $\mathcal{O} _{\rho_1 \ldots \rho _N}$.  For the intrinsic-parity-even Lagrangian at order $p^8$~\cite{Bijnens:2018lez} it was shown that this relation does not contribute, since any monomial at most has 4 independent Lorentz indices. For the intrinsic-parity-odd Lagrangian, the Schouten identity does contribute already from order $p^6$ \cite{Bijnens:2001bb,Ebertshauser:2001nj}.

In $d=4$ it is sufficient to consider 5 indices in a certain operator at a time to apply Eq.~(\ref{eq:schoutenid}). We thus need to consider $\mathcal{O} _{\rho_1 \rho_2 \rho_3 \rho_4 \rho _5}$. With an overall Levi-Civita tensor $\epsilon ^{\mu _i \mu _j \mu _k \mu _l}$ in each operator and since there at most are two different internal indices $\nu_1$ and $\nu _2$ in a given operator, there are two distinct cases:
\begin{align}
\textrm{\underline{Case 4+1:}}& \quad \mathcal{O} _{\mu _i \mu _j \mu_k \mu_l \nu _m} \, ,
\\
\textrm{\underline{Case 3+2:}}& \quad \mathcal{O} _{\mu _i \mu _j \mu_k \nu_l \nu _m} \, .
\end{align}
These two classes of operators give two sets of relations in Eq.~\eqref{eq:schoutenid}.
The 4+1 type reduces Eq.~\eqref{eq:schoutenid} from 120 to 5 terms using the symmetry of the Levita-Civita tensor and similarly the 3+2 type reduces to 20 terms. It turns out that the 20 terms of the 3+2 cases can be written in terms of six different combinations of the 4+1 terms. The 3+2 case thus does not lead to any new relations.
It is sufficient to consider the five term relation
\begin{align}\label{eq:schoutenidfinal}
    0= \sum _{\{\sigma (\mu_1 \mu _2 \mu _3 \mu _4 \nu _i)\}} \pm\mathcal{O} _{\sigma(\mu_1 \mu _2 \mu _3 \mu _4 \nu _i)} \, ,
\end{align}
where $\nu _i$ can be either $\nu _1$ or $\nu_2$. To generate these relations, we take each operator in the over-complete basis and anti-symmetrize over $\mu_1, \mu _2, \mu _3 , \mu _4$ and $\nu _i$ according to Eq.~(\ref{eq:schoutenidfinal}).

\subsection{Additional relation with $f_{-}^{\mu \nu}$}
 There is an additional relation at general $N_f$. In the absence of external fields it states that $\nabla _\mu u_\nu$ is symmetric in the two indices $\mu$ and $\nu$, but the general relation is
\begin{align}
\label{eq:umunufmunu}
\nabla _{\mu} u_{\nu}-\nabla _{\nu} u_{\mu}+f_{-\mu \nu} = 0 \, .
\end{align}
This relation is implemented by looking for all occurrences of $\nabla _{\mu} u_{\nu}$ in the operators and replacing each with the relation in Eq.~(\ref{eq:umunufmunu}).   

\subsection{Cayley-Hamilton relations at fixed $N_f$}
The Cayley-Hamilton theorem states that any $N_f\times N_f $ matrix $A$ satisfies its own characteristic equation, $p(A) =0$~\cite{Bijnens:2018lez}. Since the characteristic equation is a polynomial equation of order $N_f$, the relations are of varying complexity and lead to very different numbers of monomials in the final Lagrangians. Here we quote the results as given in Ref.~\cite{Bijnens:2018lez}. For two flavours and $2\times 2$ matrices $A$, $B$ and $C$ such that $A=B+C$, one has the relation
\begin{align}
\label{eq:CH2}
 N_{f}=2: \; \; \; \; & \left\{ B,C\right\} -B \langle C \rangle - C \langle B \rangle -\langle BC\rangle +\langle B \rangle \langle C\rangle =0 \, . 
\end{align}
For three flavours and $3\times 3$ matrices $A$, $B$, $C$ and $D$ such that $A=B+C+D$, the relation is instead
\begin{align}
\label{eq:CH3}
 N_{f}=3: \; \; \; \; & BCD +DBC +CBD+DCB +CDB +BDC  -DB \langle C\rangle
 \nonumber \\
 &-BD \langle C\rangle   -BC \langle D\rangle  - CB \langle D\rangle - DC \langle B\rangle - CD \langle B\rangle -D\langle BC \rangle 
\nonumber \\ 
& - B\langle CD\rangle -C \langle B D \rangle - \langle BCD \rangle - \langle CBD\rangle + D \langle B\rangle \langle C\rangle + B\langle C\rangle \langle D\rangle 
\nonumber \\
&+ C\langle B\rangle \langle D\rangle + \langle D \rangle \langle BC\rangle +\langle B\rangle \langle CD \rangle +\langle C \rangle \langle BD \rangle - \langle B \rangle \langle C\rangle \langle D\rangle = 0 \, .
 \end{align}
 
 It is important to take into account that the matrices $B$, $C$ and $D$ can be any combinations of building blocks in a given operator. For example, let us consider the two-flavour case and the operator
 \begin{align}
\mathcal{O} = \trace{ \chi _{-} u_{\mu } u _{\nu} \nabla _{\alpha} u_{\rho} \nabla ^{\alpha} u_{\sigma} } \epsilon^{\mu \nu \rho \sigma} \, . 
 \end{align}
 Inside the trace there are 5 different building blocks that could be identified as either $B$ or $C$ in Eq.~(\ref{eq:CH2}). These could be $B=\chi _{-}$, $C = u_{\mu }$, or $ B= u_{\mu }u _{\nu}$ and $C= \nabla _{\alpha} u_{\rho} \nabla ^{\alpha} u_{\sigma}$ and so on. For three active flavours one would need $B$, $C$ and $D$ in Eq.~(\ref{eq:CH3}). We generate Cayley-Hamilton relations by identifying all such possible matrix clusterings in the operators of the over-complete basis.

 We have also used the Cayley-Hamilton relation for $N_f=4,5$ but they are too long to reproduce here. Instead we simply provide the number of terms also for these cases. For $N_f=6$, there are no Cayley-Hamilton relations at all for our operators.

\section{The anomalous chiral Lagrangian at order $p^8$}\label{sec:lagresult}
Having obtained all operator relations in the over-complete basis, a minimal Lagrangian basis can be determined.  In this section we outline our approach to obtain the basis. The main idea is to construct a matrix of the $N_{ \textrm{rel}}$ relations found in section~\ref{sec:relations}, which can be solved to remove redundant monomials in the over-complete set with $N_{8}^{0}$. If we collect the $N_{8}^{0}$ monomials into a vector with components $\mathcal{O}_{i}^{(8)}$ and the coefficients in the linear relations in a relation matrix, $R$, of size $N_{ \textrm{rel}} \times N_{8}^{0}$, we may write the set of relations on the simple form
\begin{align} \label{eq:relationmatrix}
0 = \sum _{j=1}^{N_{8}^{0}} R_{ij} \, \mathcal{O}_{j}^{(8)} \, .
\end{align}
Many of the relations are linearly dependent, which means that the operator basis can be reduced to $N_8<N_{8}^{0}$ monomials (cf.~Eq.~(\ref{eq:lagdef2n})). The number of linearly independent rows of the relation matrix is the rank of the matrix, $\textrm{rnk} \, R $, from which we immediately find 
\begin{align}
    N_8 = N_{8}^{0}- \textrm{rnk} \, R \, .
\end{align}

\subsection{Reduction to a minimal basis} \label{sec:reduction}

To save computational resources and time, some trivial relations between monomials are employed at the beginning of the calculation to obtain the initial basis. In particular, redundant terms resulting from the cyclicity of traces and (anti)symmetry of indices were removed, as well as vanishing terms due to $\trace{u_\mu} = 0$ and $\trace{f_{\pm\, \mu \nu}}=0$. The latter requirement is relaxed to just $\trace{f_{-\, \mu\nu}} = 0$ for $N_f=2$ since, in the anomalous sector, one must include a (non-invariant) singlet-vector source~\cite{Bijnens:2001bb}. Nevertheless, below we present the size of the minimal basis for both scenarios across all studied $N_f$. All bases are presented in the supplementary material due to their length.

At order $p^8$, this initially over-complete basis contains $N^0_8 = 31,568$ monomials for $\trace{f_{\pm\, \mu \nu}}=0$, and 36,196 including  the singlet trace. By employing the identities outlined in section~\ref{sec:relations} we find about $N_\text{rel} \approx 140,000$ linear combinations of monomials that vanish for $N_f=2$, 75,000 for $N_f=3$ and $65,000$ for general $N_f$. The generated relations are obviously not linearly independent among themselves. The minimal number of linearly-independent relations is determined via a process of Gaussian elimination. Each of these relations will eliminate one monomial until all connections between the terms are exhausted, resulting in a minimal Lagrangian basis.

There are, however, multiple equivalent choices of minimal bases. The preference for retaining specific operators can be configured by reordering them before the Gaussian elimination process, with those appearing earlier in the basis being removed before other equivalent choices.

To facilitate the reproduction of our results, the criteria to reorder the operators are enumerated in the following, by order of importance. By type of field in the monomial, the following criteria are enforced: 
\begin{enumerate}
	\item First, we remove terms with only $u_\mu$ such that they are as few as possible in the final basis. This criterion allows to compare the number of operators to other approaches \cite{Kampf:2021jvf,Low:2022iim} that do not consider the external fields.
	\item Then, terms with $\chi_{\pm}$ and no $f_{\pm \, \mu\nu}$ are preferentially removed.
	\item Then, terms with $f_{\pm \, \mu\nu}$ and no $\chi_{\pm}$ are preferentially removed. 
\end{enumerate}

Within each category, the preference for retention of monomials for the final basis is as follows:
\begin{enumerate}
	\item[5.] Operators with fewer flavour traces are prioritized to explicitly account for large-$N_c$ counting, where each trace introduces a factor $1/N_c$,
	\item[6.] Operators with larger number of $\chi_{\pm}$,
	\item[7.] Operators with larger number of $f_{\pm \, \mu\nu}$,
	\item[8.] Operators with larger number of $u_\mu$,
	\item[9.] Operators with a smaller number of derivatives,
	\item[10.] Operators with a smaller number of sequential derivatives, on the same object.
\end{enumerate}

Technically, operators are assigned a score and reordered according to it. In this score, every digit represents one of the specified criteria, arranged in order of importance. The score of each term is nevertheless not unique as there are operators with the same building blocks that differ in their order, or in the position of indices, or have the derivatives acting on different objects. Consequently, it is not possible to entirely reproduce the final basis from these criteria. It is possible, however, to reproduce the number of terms in each block of the Lagrangian with a specific combination of fields and derivatives and, in any case, the equivalence of two bases can always be verified through a process of Gaussian elimination. For reference, only 205 of the 999 monomials in the final basis for general $N_f$ and $\trace{f_{\pm\, \mu \nu}}=0$ are identical in our two independent calculations. They are, however, equivalent bases and do agree on the numbers in Tables~\ref{tab:resultsNoTr} and \ref{tab:resultsTr}.

In the code of the calculation, time efficiency is crucial for managing these large systems of equations. Therefore, the decomposition of the operator-relation matrix is implemented using C++ code. In the {\sc Form} version of our calculation, we use a private implementation of Gaussian elimination with exact arithmetic via the GNU multiple precision library (GMP)~\cite{Granlund:2016}. In the \textsc{Mathematica} version of our code, \texttt{MINIBAR}~\cite{MINIBAR} includes an interface to the matrix decomposition methods of SuiteSparseQR~\cite{SPQR}, based on C++, which are optimized for sparse matrices.

\subsection{Contact terms }\label{sec:contact}

The contact terms, i.e.~those only involving external fields, are considered separately in our calculation. To make the presence of these operators explicit in our Lagrangian, the LR basis of building blocks in Eq.~\eqref{eq:defU} is employed to find all invariant structures and an analogous procedure is followed to find a minimal set of linearly independent monomials. In this case, the relevant identities to find operator relations are the partial integration relations, the Schouten identity, the Cayley-Hamilton identities, the Bianchi identity and the commutation of derivatives. Some of these take a different shape in the LR basis, as the different chiral transformation properties of the involved building blocks must be taken into account.

The Bianchi identities are written in this case as
\begin{align}
D_\mu F_{L\nu\rho}+D_\nu F_{L\rho\mu}+D_\rho F_{L\mu\nu}&\,=0\,,
\nonumber\\
D_\mu F_{R\nu\rho}+D_\nu F_{R\rho\mu}+D_\rho F_{R\mu\nu}&\,=0\,.
\end{align}
The commutation of covariant derivatives depends on the chiral transformation of the matrix $O$ on which they act and reads
\begin{align}\label{eq:fsdeflr}
D_\mu D_\nu O -D_\nu D_\mu O+iF_{R\mu\nu}O-iOF_{L\mu\nu}=0,&\quad
O\longrightarrow g_ROg_L^\dagger\,,
\nonumber\\
D_\mu D_\nu O -D_\nu D_\mu O+iF_{L\mu\nu}O-iOF_{R\mu\nu}=0,&\quad
O\longrightarrow g_L O g_R^\dagger\,,
\nonumber\\
D_\mu D_\nu O -D_\nu D_\mu O+iF_{R\mu\nu}O-iOF_{R\mu\nu}=0,&\quad
O\longrightarrow g_ROg_R^\dagger\,,
\nonumber\\
D_\mu D_\nu O -D_\nu D_\mu O+iF_{L\mu\nu}O-iOF_{L\mu\nu}=0,&\quad
O\longrightarrow g_LOg_L^\dagger\,.
\end{align} 
For the Cayley-Hamilton relations, as written in Eq.~\eqref{eq:CH2} and \eqref{eq:CH3}, one must ensure that the matrices $B$, $C$ and $D$ share the same chiral transformation properties in order to permute them.

It is important to remember that $\chi $ and $\tilde{\chi}$ by definition are not independent, namely $\tilde\chi = \left(\det(\chi)\chi^{-1}\right)^\dagger$. 
This in particular implies that $\chi \tilde{\chi}^\dagger = \det (\chi) $ which is proportional to the unit matrix. Similarly, $\trace{\chi \chi^\dagger} = \trace{\tilde{\chi} \tilde{\chi}^\dagger } $. These relations also have to be considered in addition to the ones above. As a final remark, this can be explicitly checked if we write the element of the matrix $\chi$ as $\chi_{ij} = x_{ij}$. The definition of $\tilde{\chi}$ then gives in the two-flavour case that
\begin{align}
\label{eq:chitilde}
&N_f=2:
\qquad 
 \tilde\chi \,=\begin{pmatrix}x^*_{22} & -x_{21}^*\\-x^*_{12} & x_{11}^*\end{pmatrix}\,,
\end{align}
For three flavours the same procedure can be followed, but we only need it for $N_f=2$ here.
The reason is that $\tilde\chi$ for $N_f\ge3$ is at least order $p^3$, and it must come together with $\chi^\dagger$ or $\tilde\chi^\dagger$ which is at least order $p^2$. At order $p^8$ this leaves at most order $p^3$ and thus no room for derivatives or field-strengths to match the number of indices with the overall $\epsilon^{\mu\nu\rho\sigma}$.

Finally, we obtain the minimal basis of monomials with only external fields. In the case of $\trace{f_{\pm\, \mu\nu}} = 0$, which in the LR basis reads $\trace{F_{L\, \mu\nu}} = \trace{F_{R\, \mu\nu}} = 0$, we find no contact terms for the cases $N_f=3$ and general $N_f$, and two monomials are obtained for $N_f=2$. These are 
\begin{align}
    N_f=2:& 
    \left\{
    \begin{array}{l}
        i   \trace{  \chi  F _{L \mu \nu} \tilde{\chi } ^{\dagger} F _{R \rho \sigma} } -i   \trace{  \chi  ^{\dagger} F _{R \mu \nu} \tilde{\chi }  F _{L \rho \sigma} }   \, ,     
        \\ \\
        i \left[ \trace{  \chi  \tilde{\chi } ^{\dagger} }  -   \trace{  \chi  ^{\dagger} \tilde{\chi }  } 
\right]  \left[ \trace{  F _{L \mu \nu} F _{L \rho \sigma} } +\trace{  F _{R \mu \nu} F _{R \rho \sigma} } \right]     \, .
    \end{array}
    \right. 
\end{align}

Including the singlet vector source, the relations $\trace{f_{-\, \mu\nu}} = 0$ and $\trace{f_{+\, \mu\nu}} \neq 0$ imply that, in general, $\trace{F_{L\, \mu\nu}} = \trace{F_{R\, \mu\nu}} \neq 0$. In this case, the monomials are
\begin{align}
    N_f , \text{ and } N_f=3:& 
    \left\{
    \begin{array}{l}
       \trace{  D ^{\beta} F _{L \rho \sigma}+ D ^{\beta} F _{R \rho \sigma} }
\left[ \trace{  F _{L ~\mu} ^{\alpha} D _{\alpha} F _{L \beta \nu} }  
- \trace{  F _{R ~\mu} ^{\alpha} D _{\alpha} F _{R \beta \nu} } \right]
\, ,
\\
           \\
           \trace{  D ^{\beta} F _{L \rho \sigma}+ D ^{\beta} F _{R \rho \sigma} } 
\left[  \trace{  F _{L ~\beta} ^{\alpha} D _{\alpha} F _{L \mu \nu} }  
- \trace{  F _{R ~\beta} ^{\alpha} D _{\alpha} F _{R \mu \nu} } \right] \, ,
           \\   
    \end{array} 
    \right.
    \\
    ~ &  \nonumber \\
    N_f=2:& 
    \left\{
    \begin{array}{l} \label{eq:contactNf2}
      i   \trace{  \chi  F _{L \mu \nu} \tilde{\chi } ^{\dagger} F _{R \rho \sigma} } -i   \trace{  \chi  ^{\dagger} F _{R \mu \nu} \tilde{\chi }  F _{L \rho \sigma} }
      \, ,
      \\
      \\
         i\left[   \trace{  \chi  \tilde{\chi } ^{\dagger} } 
 -   \trace{  \chi  ^{\dagger} \tilde{\chi }  }
\right]
\left[ \trace{  F _{L \mu \nu} F _{L \rho \sigma} } 
+\trace{  F _{R \mu \nu} F _{R \rho \sigma} } 
\right]
\, ,
       \\
       \\
       i \trace{  F _{L \mu \sigma} + F _{R \mu \sigma} }  \trace{  F _{L \nu \rho} + F _{R \nu \rho} } \left[ 
   \trace{  \chi  \tilde{\chi } ^{\dagger} }  
-  \trace{  \chi  ^{\dagger} \tilde{\chi }  } \right] 
\, ,
\\
\\
 \trace{  D ^{\beta} F _{L \rho \sigma}+ D ^{\beta} F _{R \rho \sigma} }
\left[ \trace{  F _{L ~\mu} ^{\alpha} D _{\alpha} F _{L \beta \nu} }  
- \trace{  F _{R ~\mu} ^{\alpha} D _{\alpha} F _{R \beta \nu} } \right]
\, ,
\\
           \\
           \trace{  D ^{\beta} F _{L \rho \sigma}+ D ^{\beta} F _{R \rho \sigma} } 
\left[  \trace{  F _{L ~\beta} ^{\alpha} D _{\alpha} F _{L \mu \nu} }  
- \trace{  F _{R ~\beta} ^{\alpha} D _{\alpha} F _{R \mu \nu} } \right] \, .
    \end{array} 
    \right.
\end{align}
Therefore, in the physical cases for the anomalous chiral Lagrangian at order $p^8$ we found no contact terms for $N_f$ and $N_f=3$, and 5 monomials for $N_f=2$, given in Eq.~\eqref{eq:contactNf2}. For comparison, at order $p^6$ there were no contact terms for any $N_f$, regardless of the inclusion of the singlet vector source for $N_f=2$~\cite{Bijnens:2001bb}.

Once the minimal basis of contact terms is obtained, they must be added to the main Lagrangian, written in the $u$ basis. However, to keep a genuinely minimal Lagrangian basis, the same number of monomials must be removed from the general Lagrangian. Technically, we achieve this by repeating the Gaussian elimination procedure for the over-complete basis of $N^0_8$ monomials, this time extended with the contact terms at the end. Accordingly, the relation matrix $R_{ij}$ in Eq.~(\ref{eq:relationmatrix}), is expanded with one relation per contact term, in which this is set equal to its translation into the $u$ basis, following section~\ref{sec:translation}. This procedure also serves as a consistency test, confirming that the basis of contact operators, derived in the LR parametrization of the fields, remains minimal under the set of operator relations obtained using the $u$ parametrization for the general Lagrangian.

\subsection{The Lagrangian}
We now present our results in Tables~\ref{tab:resultsNoTr}--\ref{tab:resultsTr}, where we give the number of monomials for varying numbers of flavours and with certain combinations of building blocks. The Lagrangian bases for $N_f$, $N_f=3$ and $N_f=2$ are presented in the supplementary material. Table~\ref{tab:resultsNoTr} corresponds to the case $\langle f_{+}^{\mu \nu}\rangle = 0$. Note that for two flavours this excludes the singlet vector source, which corresponds to the case studied in Ref.~\cite{Graf:2020yxt}. The numbers on the first row where all fields are included agree with Table 10 of Ref.~\cite{Graf:2020yxt}. At order $p^6$ when all fields are included, there are 24 monomials for a general $N_f$, 23 for $N_f=3$ and 5 for $N_f=2$~\cite{Ebertshauser:2001nj,Bijnens:2001bb}. Comparing to Table~\ref{tab:resultsNoTr} we see a sizeable increase in the numbers, and a bigger difference between $N_f$ and $N_f=3$. 

In Table~\ref{tab:resultsTr} we include the singlet vector source $\langle f_{+}^{\mu \nu}\rangle \neq 0$, which has not been considered before in the literature. We have for completeness included it for all $N_f$, even though it in reality only matters for $N_f=2$. We observe a mild increase in the number of monomials compared to Table~\ref{tab:resultsNoTr}. The two last rows are identical for the two tables since the field $f_+^{\mu \nu}$ is not included there. The number of monomials for order $p^6$ when all fields including the singlet source is 13~\cite{Ebertshauser:2001nj,Bijnens:2001bb}.

\begin{table}[t]
\begin{center}  
\renewcommand{\arraystretch}{1.2}
\begin{tabular}{|c|c|c|c|c|c|c|}
\hline
$\trace{f_{+}^{\mu \nu}}=0 $   & $N_{f}$ & $N_{f}=5$ & $N_{f}=4$ & $N_{f}=3$  & $N_{f}=2$  \\ \hline 
Full                    &  999 & 998 & 950 & 705 &  92  
\\ \hline
No $\chi _{\pm}$        &  565 & 564 & 525 & 369 & 0   
\\ \hline
No $f _{\pm}^{\mu \nu}$ &  79 & 79 & 73 & 45  &  2  
\\ \hline
Only $u_\mu$            &  36 & 36 & 31 & 16 &  0    
\\ \hline
    \end{tabular}
   \renewcommand{\arraystretch}{1}
    \caption{Number of monomials in the obtained minimal basis. We here also include the cases where only a subset of the external field building blocks contribute. For these numbers we used $\trace{f_{+}^{\mu \nu}}=0$, which means that the singlet non-zero trace physically relevant for $N_f = 2$ is not included.}\label{tab:resultsNoTr}
  \end{center}
\end{table}

\begin{table}[t]
\begin{center}  
\renewcommand{\arraystretch}{1.2}
\begin{tabular}{|c|c|c|c|c|c|c|}
\hline
$\trace{f_{+}^{\mu \nu}}\neq 0$   & $N_{f}$ & $N_{f}=5$ & $N_{f}=4$  & $N_{f}=3$  & $N_{f}=2$  \\ \hline 
Full                    &  1210 & 1209 & 1161 & 892 & 211 
\\ \hline
No $\chi _{\pm}$        &  702  & 701 & 662 & 486 & 77    
\\ \hline
No $f _{\pm}^{\mu \nu}$ &  79  & 79 & 73 &  45 & 2   
\\ \hline
Only $u_\mu$            &  36 & 36 & 31  &  16 & 0     
\\ \hline
    \end{tabular}
   \renewcommand{\arraystretch}{1}
    \caption{Number of monomials in the obtained minimal basis when $\trace{f_{+}^{\mu \nu}}\neq 0$. Again we include the cases where only a subset of the external field building blocks contribute.}\label{tab:resultsTr}
  \end{center}
\end{table}

We may next use our derived bases to connect to the literature. In Ref.~\cite{Kampf:2021jvf}, the non-linear sigma model was studied at higher orders. A particular case in this reference corresponds to the anomalous Lagrangian derived here with $N_f$ flavours, without external fields and for monomials with only one flavour trace (i.e.~in the large-$N_c$ limit, with $N_c$ colours). This means that we can fill in the missing numbers in Table 6 of Ref.~\cite{Kampf:2021jvf}, corresponding to the number of monomials with either 5 or 7 mesons in our ChPT Lagrangian. We obtain 3 terms contributing with 5 meson interactions and 15 with 7 meson interactions agreeing with the amplitude results derived in Ref.~\cite{Kampf:2021jvf}.

\section{Conclusions}\label{sec:conclusion}
In this paper we have derived an explicit basis for the mesonic chiral Lagrangian of odd intrinsic parity at order $p^8$. This completes the knowledge of the effective Lagrangian at order $p^8$ together with previous work in the even intrinsic parity sector by two of the authors~\cite{Bijnens:2018lez}. 

We have considered 2, 3 as well as a general number of light quark flavours $N_f$. 
The number of monomials in the minimal basis for the respective cases are
\begin{align}\label{eq:physres}
    \underline{N_f:} & \quad 999 \, , \nonumber \\
    \underline{N_f=3:} & \quad 705 \, , \nonumber \\
    \underline{N_f=2:} & \quad 211 \, .
\end{align}
We emphasise that for $N_f=2$ the number includes the singlet non-zero trace $\trace{f_{+}^{\mu \nu}}\neq 0$. The corresponding numbers at order $p^6$ are 24, 23 and 13~\cite{Ebertshauser:2001nj,Bijnens:2001bb}. For completeness, we have also derived the numbers with $\trace{f_{+}^{\mu \nu}}\neq 0$ for $N_f=3$ and a general $N_f$, given in section~\ref{sec:lagresult}. Our procedure is based on an explicit enumeration of all possible monomials that can be built from the underlying chiral building blocks, with possible referring to invariance under the chiral symmetry as well as Lorentz, parity, charge and hermitian conjugation transformations. Our results confirm and also extend those of Ref.~\cite{Graf:2020yxt} using an independent Hilbert series method to determine the numbers.

Some of the monomials in the bases in Eq.~(\ref{eq:physres}) are unmeasurable as they only depend on external fields. The number of contact terms for different values of $N_{f}$ are
\begin{align}
    \underline{N_f:} & \quad  0 \, , \nonumber \\
    \underline{N_f=3:} & \quad  0 \, , \nonumber \\
    \underline{N_f=2:} & \quad  5  \, .
\end{align}
At order $p^6$ there are no contact terms~\cite{Ebertshauser:2001nj,Bijnens:2001bb}. 

In the supplementary material we provide explicit bases for all the studied Lagrangians, explicitly including also the contact terms for the general $N_f$ and $N_f=2,3$ cases both in pdf and machine readable format. These bases correspond to the ones obtained with the \textsc{Mathematica} version of the calculation.

The phenomenological impact of the derived Lagrangian is at present still rather limited, mainly due to large number of low-energy constants multiplying the monomials in the minimal basis. For phenomenological applicability one would also need to renormalize the low-energy constants. So far only one full calculation to order $p^8$ exists, $\pi^0\to\gamma\gamma$~\cite{Kampf:2009tk,Bijnens:2010pa} and work on leading logarithms~\cite{Bijnens:2012hf}.
The situation is the same for the Lagrangian of even intrinsic parity~\cite{Bijnens:2018lez,Hermansson-Truedsson:2020rtj}. Despite this practical limitation for phenomenology, the results are of interest in the amplitude community, e.g.~as the Lagrangian here can be used to study the non-linear sigma model in the limit of a large number of colours~\cite{Kampf:2021jvf}.

\section*{Acknowledgements}
J.~B.~and J.~R.~V.~are supported by the Swedish
Research Council grants contract numbers 2016-05996 and 2019-03779. 
N.~H.-T.~was originally funded by the Swedish Research Council, project number 2021-06638, and now by the UK Research and Innovation, Engineering and Physical Sciences Research Council, grant number EP/X021971/1. N.~H.-T.~wishes to thank the Higgs Centre for Theoretical Physics at The University of Edinburgh for hosting him as a visitor when the first part of this work was done.


\bibliographystyle{JHEP}
\bibliography{refs}

\end{document}